\documentclass[a4paper,10pt]{article}
\usepackage[utf8]{inputenc}

\usepackage{amsmath}
\usepackage{amssymb}
\usepackage{hyperref}
\usepackage{graphicx}
\usepackage{verbatim}
\usepackage{geometry}
\def\I{\mathcal{I}}
\usepackage{tikz}
\tikzstyle{every picture}=[level distance = 8mm, baseline=-0.5ex]
\tikzstyle{prop}=[shape=circle,minimum size=6mm, draw=black!80, fill=green!30]

\begin{document}

\title{Higher Order Corrections to the Asymptotic  Perturbative Solution of a {S}chwinger--{D}yson Equation.}
\author{Marc~P.~Bellon${}^{1,2}$, Pierre~J.~Clavier${}^1$\\
\normalsize \it ${}^1$UPMC Univ Paris 06, UMR 7589, LPTHE, F-75005, Paris, France\\
\normalsize \it $^2$CNRS, UMR 7589, LPTHE, F-75005, Paris, France}

\date{}

\maketitle

\begin{abstract}
Building on our previous works on perturbative solutions to a Schwinger--Dyson for the massless Wess--Zumino model, we show how to compute \(1/n\) corrections to its asymptotic behavior. 
The coefficients are analytically determined through a sum on all the poles of the Mellin transform of the one loop diagram. We present results up to the fourth order in \(1/n\) as well as a 
comparison with numerical results. Unexpected cancellations of zetas are observed in the solution, so that no even zetas appear and the weight of the coefficients is lower than expected, which 
suggests the existence of more structure in the theory.
\end{abstract}

\noindent {\em Keywords}: Renormalization; Schwinger--Dyson equation; Borel transformation.

\noindent {\em 2010 Mathematics Subject Classification}: 81T15;81T17.

\section*{Introduction}
Quantum Field Theory (QFT) is the set of mathematical tools allowing us to describe (at least) three of the four known fundamental interactions, as well as a number of phenomena in statistical 
mechanics and solid state physics. Perturbative solutions are behind successes in particle physics (QED, electroweak theory and QCD, put together into the Standard Model).

Nonetheless, the present state of perturbative QFT is not very satisfying since the maximum computed order has grown rather slowly,  due to the very fast growth of the number of graphs to 
compute with the order in the perturbation theory, as well as rising difficulties in their evaluations. It is therefore not clear how we may know which (nor how many) graphs of a given order 
will give a significant contribution. There are also physically relevant situations where the perturbation theory breaks down, or is not efficient (e.g., low-energy QCD). Many solutions have 
been probed for these issues, like lattice QCD, effective models.

An other way to overcome them is to use Schwinger--Dyson equations. Although these equations come from perturbation theory, they can also be studied non-perturbatively. The main difficulty is 
that one must resort to truncated versions, which may not display important properties of the full theory, like the Ward identities associated to gauge invariance.  Nevertheless, important 
qualitative properties of QCD, like spontaneous chiral symmetry breaking and color confinement, can be obtained from the numerical study of simplified Schwinger--Dyson equations (see e.g., 
\cite{AdRo12} and reference therein). However, exact solutions of Schwinger--Dyson equations are only known for a linear  one~\cite{BrKr99}.

Nevertheless, much can be learned about the perturbative series of quantum field theories through the solution of a particular Schwinger--Dyson equation, in the four-dimensional supersymmetric Wess--Zumino model.
The breakthrough came from the recognition of a Hopf algebra structure on the Feynman graphs of a QFT in \cite{CoKr99}, \cite{CoKr00}. This Hopf algebra structure takes care
of the usual combinatorial technicalities of QFT and set them into an algebraic framework. A key point of this work is that the renormalisation group can be 
studied in this algebraic context. So one ends up with new relations on the quantities relevant for the study of the renormalisation group. One can plug these relations 
into the Schwinger--Dyson equation of the massless Wess--Zumino model so it becomes much simpler to solve. In particular, we do not need a huge numerical study to extract results.

The method that we are going to use here has been fully explained in \cite{Be10a} and \cite{BeSc12}. The starting point of those computations was the work \cite{KrYe2006} where it was made 
clear how the renormalisation group equation can be used to deduce the full propagator from the anomalous dimension and how the anomalous dimension of a theory can be derived from its 
Schwinger--Dyson equation.

This paper has five parts. The first is a brief review of the renormalisation group seen in the framework of Connes--Kreimer Hopf algebra of renormalisation, the 
Schwinger--Dyson equation of the massless Wess--Zumino model and the methods and results of \cite{Be10a}. In the second part we define a powerful change of variables which will drastically reduce the complexity of the 
computations to obtain subdominant terms in the higher orders of the perturbative solution. The results of \cite{Be10a} are given with the new set of variables. In the third part we write down the Schwinger--Dyson equation 
with the first contributions from every poles of the Mellin transform (which we will define later) and solve them. The fourth part is devoted to higher order computations, up to the fourth 
order. This fourth order is found to have an unexpected behaviour, as explained at the very end of this fourth part. Finally, we compare our result to the numerical results of \cite{BeSc08}. An 
excellent agreement is found and the two last unconstrained parameters of our method are numerically determined.

\section{Methods and previous results}
\label{methods}

\subsection{Renormalisation Group}

In a given massless QFT, one can expand the two point point function in power of the logarithm of the impulsion $L=\ln(p^2/\mu^2)$.
\begin{equation}
 G(L) = 1+\sum_{k=1}^{+\infty}\gamma_k\frac{L^k}{k!}.
\end{equation}
The $\gamma$'s are themselves functions of the fine structure constant of the theory, which we will denote by $a$.
\begin{equation}
 \gamma_k = \sum_{n=0}^{+\infty}\gamma_{k;n}a^n.
\end{equation}
As proven in the thesis \cite{Ye08} and in the article \cite{BeSc08} the renormalisation group yields a recursion relation on the $\gamma_k$'s:
\begin{equation} \label{recursion_gamma_vieux}
 \gamma_{k+1} = (\gamma_1+\beta a\partial_a)\gamma_k.
\end{equation}
Here $\beta$ is the $\beta$-function of the theory and $\partial_a$ the derivative with respect to $a$. This result stems from the work of Connes and Kreimer~\cite{CoKr00},
where the renormalisation group was shown to be a one parameter subgroup of the group of characters of the Hopf algebra of the Feynman graphs, as well as the existence of a sub-Hopf algebra 
generated by the coefficients of the Green functions~\cite{vSu2007,BeSc08}. Then the equation (\ref{recursion_gamma_vieux}) comes from the fact that the $\beta$-function is a derivative.

Now, in the massless Wess--Zumino model, the vertices are not divergent and therefore do not need to be renormalized. Only the propagator is affected by the renormalisation group and, thanks to 
the supersymmetry, all components of the supermultiplet get the same renormalisation factor. The Callan--Symanzik equation then leads to
 \[\beta = 3\gamma_1.\]
The proof of this result is detailed in \cite{Piguet}. Hence the recursion on the $\gamma$'s is very simple.
\begin{equation} \label{recursion_gamma}
 \gamma_{k+1} = \gamma_1(1+ 3a\partial_a)\gamma_k.
\end{equation}
In this model, all the coefficients of the expansion of $G(L)$ are therefore simple functions of the first coefficient.

\subsection{Non-linear Schwinger--Dyson equation}

In the massless Wess--Zumino model, the only important Schwinger--Dyson equation for the computation of renormalization group functions is therefore the one for the propagator and we take its first approximation, which is the simplest non-linear one and can be
graphically described by:
\begin{equation}\label{SDnlin}
\left(
\tikz \node[prop]{} child[grow=east] child[grow=west];
\right)^{-1} = 1 - a \;\;
\begin{tikzpicture}[level distance = 5mm, node distance= 10mm,baseline=(x.base)]
 \node (upnode) [style=prop]{};
 \node (downnode) [below of=upnode,style=prop]{}; 
 \draw (upnode) to[out=180,in=180]   
 	node[name=x,coordinate,midway] {} (downnode);
\draw	(x)	child[grow=west] ;
\draw (upnode) to[out=0,in=0] 
 	node[name=y,coordinate,midway] {} (downnode) ;
\draw	(y) child[grow=east]  ;
\end{tikzpicture}.
\end{equation}
This simplest equation is non trivial to solve since the  loop integral depends on the unknown propagator. However, at a given loop order, the full propagator is the free propagator times the 
two-point function, which has a finite expansion in the logarithm of the momentum.
\begin{equation}
 P(p^2/\mu^2) = \frac{1}{p^2/\mu^2}\left(1+\sum_{k=1}^{+\infty}\gamma_k\frac{L^k}{k!}\right)
\end{equation}
The next move is to take the Mellin transform of this loop integral. This is nothing but noticing:
\begin{equation}
 \left(\ln\frac{p^2}{\mu^2}\right)^k = \left.\left(\frac{\text{d}}{\text{dx}}\right)^k\left(\frac{p^2}{\mu^2}\right)^x\right|_{x=0}.
\end{equation}
Then, after an exchange between sum, derivatives and integral, one has only one integral (a Mellin transform) to perform.
\begin{align}
 \text{I}(q^2/\mu^2,x,y) & = \frac{g^2}{8\pi^4}\int\text{d}^4p\frac{1}{\left(p^2/\mu^2\right)^{1-x}[(q-p)^2/\mu^2]^{1-y}} \nonumber \\
                       & = a\left(\frac{q^2}{\mu^2}\right)^{x+y}\frac{\Gamma(-x-y)\Gamma(1+x)\Gamma(1+y)}{\Gamma(2+x+y)\Gamma(1-x)\Gamma(1-y)}
\end{align}
with $a=g^2/8\pi$ and $\Gamma$ being Euler's gamma function. We will work with the derivative of this integral, which will give \(\gamma_1\),
\begin{equation}
 H(x,y) = -\left.\frac{\partial\text{I}(q^2/\mu^2,x,y)}{\partial L}\right|_{L=0}
\end{equation}
Writing $\Big({q^2}/{\mu^2}\Big)^{x+y} = \exp((x+y)L)$ one finds:
\begin{equation} \label{def_H}
 H(x,y) = a\frac{\Gamma(1-x-y)\Gamma(1+x)\Gamma(1+y)}{\Gamma(2+x+y)\Gamma(1-x)\Gamma(1-y)},
\end{equation}
which is nicely symmetric in $x$ and $y$ and finite around the origin. To write the Schwinger--Dyson equation in a compact way, let us define the transform $\mathcal{I}:\mathbb{C}[[x;y]]\rightarrow\mathbb{C}[a]$ by:
\begin{equation}
 \mathcal{I}\left(f(x,y)\right) = \left(1+\sum_{n=1}^{+\infty}\frac{\gamma_n}{n!}\frac{\text{d}^n}{\text{dx}^n}\right)\left(1+\sum_{m=1}^{+\infty}\frac{\gamma_m}{m!}\frac{\text{d}^m}{\text{dy}^m}\right)f(x,y)\bigg|_{x=y=0}.
\end{equation}
The non-linear Schwinger--Dyson equation~(\ref{SDnlin}) then gives: 
\begin{equation} \label{SDE}
 \gamma_1 = \mathcal{I}\left(H(x,y)\right)
\end{equation}
In the following, we will call this $\gamma_1$ the anomalous dimension of the theory and denote it simply by $\gamma$. 

\subsection{Contribution of the dominant poles}
\label{contribution}

In equation (\ref{SDE}), the number of different terms contributing at a given order grows quadratically with this order, making an asymptotic analysis unpractical. The solution proposed in~\cite{Be10a} is to first approximate the complicated function $H(x,y)$ by its pole parts with a suitable extension of their residues.
The contributions of every pole to the anomalous dimension can then be computed.

The function $H(x,y)$ has poles at $x;y=-k$, $k\in\mathbb{N}^*$ (we call these poles the simple ones) and at the lines $x+y=+k$, 
$k\in\mathbb{N}$ (the general poles). The simple poles are linked with the IR divergences of the loop integral while the general poles come from its UV divergences. Both kinds of pole
arise when, in the Mellin transform, a subgraph becomes scale invariant~\cite{BeSc12}. 

Now,  the simple poles can be expanded as:
\begin{equation}
 \frac{1}{k+x} = \frac{1}{k}\sum_{n=0}^{+\infty}\left(-\frac{x}{k}\right)^n,
\end{equation}
so that by~(\ref{SDE}), the contribution of such a pole to the Mellin transform is:
\begin{equation} \label{form_F}
 F_k = \frac{1}{k}\sum_{n=0}^{+\infty}\left(-\frac{1}{k}\right)^n\gamma_n
\end{equation}
with the  convention $\gamma_0=1$. Using the recurrence relation~(\ref{recursion_gamma_vieux}) between 
\(\gamma_n\) and \(\gamma_{n+1}\), one obtains:
\begin{equation} \label{equa_F}
	(\gamma + \beta a\partial_a) F_k = -k F_k + 1.
\end{equation}

For the other poles, the situation is slightly more subtle. The full proof is given in \cite{Be10a} but basically the numerators, i.e., the residues of $H(x,y)$ at those poles, must be taken in 
account from the start. We will show later that the residues of $H(x,y)$ are polynomials. Let us call $Q_k(x,y)$ the residue of $H(x,y)$  at $x+y=k$. Then the numerator at this pole is:
\begin{equation}
 N_k(\partial_{L_1},\partial_{L_2}) = Q_k(\partial_{L_1}\partial_{L_2}).
\end{equation}
The contribution $L_k$ of this pole to the Mellin transform cannot be expressed simply in terms of the \(\gamma_n\) as in ({\ref{form_F}), but it can be computed from the following equation 
that it obeys:
\begin{equation} \label{equa_H}
 (k-2\gamma - \beta a\partial_a)L_k = N_k(\partial_{L_1},\partial_{L_2})G(L_1)G(L_2)|_{L_1=L_2=0}.
\end{equation}

In~\cite{Be10a}, the function $H(x,y)$ was approximated by its first poles at $x=-1$, $y=-1$ and $x+y=+1$, giving the following approximating function in~(\ref{SDE}):
\begin{equation} \label{appr}
 \frac 1a h(x,y) = (1+xy)\left(\frac{1}{1+x} + \frac{1}{1+y}-1\right) + \frac{1}{2}\frac{xy}{1-x-y} + \frac{1}{2}xy.
\end{equation}
This means that we only use the contributions $F\equiv F_1$ of the poles $1/(1+x)$ and $1/(1+y)$ and $L\equiv L_1$ of the pole $xy/(1-x-y)$ to compute \(\gamma\). Then the 
renormalisation group equations (\ref{equa_F}) and~(\ref{equa_H}) for $F$ and $L$ and the Schwinger--Dyson equation~(\ref{SDE}) with the approximate function $h(x,y)$ defined in
 (\ref{appr}), gives the three coupled non-linear differential equations:
\begin{equation} \label{SDE_simples}
 \begin{cases}
  & F = 1 - \gamma(3a\partial_a+1)F , \\
  & L = \gamma^2 + \gamma(3a\partial_a+2)L  ,\\
  & \gamma = 2a F -a -2a\gamma( F-1) + \frac{1}{2}a(L-\gamma^2) .
 \end{cases}
\end{equation}
We look for a perturbative solution of these equations,  and expand $F$, $L$ and $\gamma$ in powers of $a$: $F=\sum f_na^n$, $L=\sum l_na^n$ and $\gamma=\sum c_na^n$. 
With the assumption that the $\{f_n\}$, the $\{l_n\}$ and the $\{c_n\}$ have a fast growth and keeping only the dominant contributions, one 
obtains three coupled recursions formulas, which can be solved to obtain the two dominant terms in each series:
\begin{align} \label{result_simple}
 \begin{cases}
  & f_{n+1} \simeq -(3n+5) f_n, \\
  & l_{n+1} \simeq 3n l_n ,\\
  & c_{n+1} \simeq -(3n+2) c_n.
 \end{cases}
\end{align}
We will see later that this results are the base for an ansatz for a systematic improvement of these asymptotic results while taking into account every pole of $H(x,y)$.

For now, let us just say that the results (\ref{result_simple}) nicely fit with the numerical study of \cite{BeSc08}. 

\section{Change of variables and first order results}\label{FirstOrder}

Now, one would like compute the $1/n$ corrections to the asymptotic solution (\ref{result_simple}). However this kind of computation turns out to be quickly tedious. To simplify our calculations, 
we separate the alternating contributions to $(c_n)$ from the ones with a constant sign. We will define two symbols for this, one to encode the asymptotic behavior coming from $l_n$ and the other one for the asymptotic behavior coming from $f_n$. They will be defined through two series:
\begin{align} \label{eq_serie_symboles}
 \begin{cases}
  & A_{n+1} = -(3n+5)A_n \\
  & B_{n+1} = 3nB_n
 \end{cases}
\end{align}
but we will only use the following two symbols corresponding to the formal series:
\begin{align}
 \begin{cases}
  & A = \sum A_n a^n \\
  & B = \sum B_n a^n
 \end{cases}
\end{align}
The relations (\ref{eq_serie_symboles}) can be expressed as differential equations on the symbols $A$ and $B$:
\begin{align} \label{relation_symboles}
 \begin{cases}
  & 3a^2\partial_a A = -A-5aA \\
  & 3a^2\partial_a B = B
 \end{cases}
\end{align}
In fact, (\ref{eq_serie_symboles}) do not entirely determine $A$ and $B$. We must have an initial condition at an order $n_0$. Then (\ref{relation_symboles}) is true up to a term of degree $n_0$. Therefore, $n_0$ shall be chosen higher than the degree to which we will compute the gamma function.

Although these relations are not totally exact they will drastically reduce the complexity of the computation of the corrections to the asymptotic behavior, allowing us to go up to the fourth order,
with computations of the same degree of complexity that those of the second order with the previous method. To do that, we will define nine unknown functions, which are the  
coefficients of \(A\) and \(B\) for the $F$, $L$ and $\gamma$ functions.
\begin{align} \label{redefinition}
 \begin{cases}
  & F = f +Ag+Bh \\
  & L = l+Am+Bn \\
  & \gamma = a(c+Ad+Be)
 \end{cases}
\end{align}
The $a$ in the ansatz for $\gamma$ comes from the $a$ in the function $H(x,y)$. In what follows, $c=c(a)$ will be called the low order part of $\gamma$.

Now, one can rewrite the Schwinger--Dyson equations (\ref{SDE_simples}) in the language of this new set of functions, with the derivatives of the $A$ and $B$ symbols
 being removed, thanks to the relations~(\ref{relation_symboles}). The new equations are found by saying that the coefficient of the symbol~$A$ and the one of the 
symbol~$B$ shall independently vanish, since the two divergences of the Mellin transform are of different nature and thereof do not talk to each other. Similarly,
the term without any symbol should also independently vanish.
Hence, this change of variables allows to efficiently separate the alternating part and the part of constant sign of the Mellin transform. It will drastically 
simplify the equations to solve since we will have three equations for each of the previous ones, with the overall system being of the same complexity than the one 
with the old formalism.

One then ends up with the nine equations:
\begin{subequations}
\begin{align}
 & f+c(3a^2\partial_a+a)f=1 \label{eq_debut} \\
 & g+d(3a^2\partial_a+a)f+c(-1-4a+3a^2\partial_a)g=0 \label{eq_g}\\
 & h+e(3a^2\partial_a+a)f+c(1+a+3a^2\partial_a)h=0 \\
 l& =a^2c^2+ac(3a\partial_a+2)l \\
 m& =2a^2dc+c(3a^2\partial_a-1-3a)m+ad(3a\partial_a+2)l \\
 n& =2a^2ec+c(3a^2\partial_a+1+2a)n+ae(3a\partial_a+2)l \label{eq_n}\\
 c& =2f-1-2ac(f-1)-\frac{1}{2}(l+a^2c^2) \\
 d& =2g+\frac{1}{2}am+2ad(1-f)-ac(2g+ad) \\
 e& =2h+\frac{1}{2}an+2ae(1-f)-ae(2f+ac) \label{eq_fin}
\end{align}
\end{subequations}
There are obviously more terms into the expansion of the equations (\ref{result_simple}), proportional to  $A^2$, $AB$,~$B^2$, but they will not be considered: if \(A\) and \(B\) begin by a large number of vanishing coefficients,  they correspond to corrections of very high order.

The perturbative solution of these equations goes as follows. We write:
\begin{equation}
 c(a)=\sum_{n=0}^{+\infty}c_na^n,
\end{equation}
and similarly for all the other functions. At each order, the equations should be solved in the right order: one shall first 
solve the equations for $f$ and $l$, then for $c$, then for $g$, $m$ and $h$, $n$, and finally for $d$ and $e$. Following this procedure, one ends up with the solution up to the 
order $a^1$.
\begin{align*}
  & f(a) = 1-a\\
  & g(a) = g_0+g_1a\\
  & h(a) = -\frac{1}{4}an_0\\
  & l(a) = 0\\
  & m(a) = 0\\
  & n(a) = n_0+n_1a\\
  & c(a) = 1-2a\\
  & d(a) = 2g_0+(-2g_0+2g_1)a\\
  & e(a) = \frac{1}{2}n_0+\frac{1}{2}(n_1-n_0)a  
\end{align*}
Here the assumption of fast growth of the series which was done in \cite{Be10a} is not necessary since the symbols $A$ and $B$ take care of the necessary properties.

The coefficients $g_1$ and $n_1$ are not specified at this stage. It is a general feature of this parametrization that one needs to go at the order $a^{p+1}$ to 
fix the parameters $g_p$ and $n_p$. Indeed, since $c_0=1$, the $a^n$ order in the equation (\ref{eq_g}) is:
\begin{equation*}
 g_n + (...)-c_0g_n + (...) = 0.
\end{equation*}
therefore does not depend on $g_n$, with a similar phenomenon appearing in (\ref{eq_n}). However,   the next order of equations (\ref{eq_g}) and~(\ref{eq_n}) is not hard and does not involve 
higher coefficients of \(d\) or \(e\), so that we obtain the solution (up to the order $a^1$) to the equations (\ref{SDE_simples}) with only two unconstrained parameters. We however need the
values of the next order 
for \(c\).
\begin{subequations}
 \begin{align}
  & f(a) = 1-a\\
  & g(a) = g_0\left(1+\frac{16}{3}a\right)\\
  & h(a) = -\frac{1}{4}an_0\\
  & l(a) = 0\\
  & m(a) = 0\\
  & n(a) = n_0\left(1-\frac{11}{3}a\right)\\
  & c(a) = 1-2a\\
  & d(a) = 2g_0\left(1+\frac{13}{3}a\right)\\
  & e(a) = \frac{1}{2}n_0\left(1-\frac{14}{3}a\right)
 \end{align}
\end{subequations}

The fact that there remain two unconstrained parameters, $g_0$ and $n_0$, is not really surprising since they  were already present in the former formalism, where the asymptotic behavior was 
inferred from the ratio of successive coefficients of the Taylor series. Since only ratios could be computed, the overall factors in the asymptotic behavior of the series for \(F\) and \(L\) 
are unconstrained. In this new formalism, equations stemming from the part linear in \(A\) are linear in the coefficients \(d\), \(g\) and~\(m\) of \(A\)  in the unknown functions: if there is 
any non trivial solution, all its multiples are also solutions. If we had not used the analysis of the previous work~\cite{Be10a}, the precise recurrence for \(A\) could have be obtained from 
the requirement of the existence of a non trivial solution. Analogous statements hold for the terms proportional to \(B\).

Up to now, we have only obtained the first two coefficients of every functions, since the other poles of the $ H(x,y)$ function will contribute to the next terms. This is the subject of the next parts.

\section{Higher poles of the Mellin transform}

\subsection{Methodology for an analysis on every pole}

To go further we must include the contributions of all the poles of the Mellin transform $H(x,y)$ to the $\gamma$ function. The equations for $F$ and 
$L$ do not change. Then we have to compute the residues of $H(x,y)$ at its various poles. They are given for $k\ge2$ by
\begin{align*}
 & \text{Res}(H,x=-k) = \frac{-y}{k-1}\prod_{i=1}^k\left(1-\frac{y}{i}\right)\prod_{i=1}^{k-2}\left(1-\frac{y}{i}\right) ,\\
 & \text{Res}(H,x=k-y) = \frac{(k-y)y}{k(k+1)}\prod_{i=1}^{k-1}\left(1-\frac{(k-y)y}{i(k-i)}\right),
\end{align*}
with the convention $\prod_{i=1}^{k-2}=1$ for $k=2$. In order to simplify the computations, we use the fact that the first polynomial is defined at $x=-k$ and the 
second at $x=k-y$ to make the numerators symmetric in $x$ and $y$. One gets:
\begin{align}
 & \text{Res}(H,x=-k) = \frac{xy}{k(k-1)}\prod_{i=1}^k\left(1+\frac{xy}{ki}\right)\prod_{i=1}^{k-2}\left(1+\frac{xy}{ki}\right) = P_k(xy) \label{residusPk}\\
 & \text{Res}(H,x=k-y) = \frac{xy}{k(k+1)}\prod_{i=1}^{k-1}\left(1-\frac{xy}{i(k-i)}\right) = Q_k(xy) \label{residusQk}.
\end{align}
The coefficients of those polynomials will be of interest. Hence we define them in the following way:
\begin{align*}
 & P_k(X) = \sum_{n=1}^{2k-1}p_{k,n}X^n \\
 & Q_k(X) = \sum_{n=1}^{k}q_{k,n}X^n.
\end{align*}
Notice that the residues at the poles in $y=-k$ are exactly the same since $H(x,y)$ is symmetric under the exchange of $x$ and~$y$. We therefore write:
\begin{eqnarray} \label{Mellin_tsfo_complete}
 \frac{1}{a}H(x,y) &=& (1+xy)\left(\frac{1}{1+x} + \frac{1}{1+y}-1\right) + \frac{1}{2}\frac{xy}{1-x-y} \nonumber\\
  &&+\sum_{k=2}^{+\infty}\left(\frac{1}{k+x} + \frac{1}{k+y} -\frac{1}{k}\right)P_k(xy) + \sum_{k=2}^{+\infty}\frac{Q_k(xy)}{k-x-y} + \tilde{H}(x,y).
\end{eqnarray}
The \(-1/k\) term coming with the poles at \(x=-k\) and \(y=-k\) does not contribute to the singularities but appears necessary to obtain the exact Taylor expansion of \(H(x,y)\) around the origin. Moreover, $\tilde{H}(x,y)$ shall be a holomorphic function and is the difference between $H(x,y)$ such as written in (\ref{def_H}) and the above expension of sums over the poles. We have checked that $\tilde{H}(x,y)$ shall be of degree at least 10. We have also verified that some infinite families of derivatives of $\tilde{H}(x,y)$ vanish at the origin. Hence we will make the conjecture $\tilde{H}(x,y)=0$ in the following, which is the raison d'\^etre of the $-1/k$. We are fairly confident that this conjecture is true, but it would be pleasant to have a proof of it, which would give a stronger ground to our computations.

To obtain the anomalous dimension of the theory at a given order~$p$, one must include additional terms of the Schwinger--Dyson equations to deal with all the contributions at this order. Indeed, we have seen in section~\ref{contribution} that the equations for $L_k$ and $\gamma$ depend on the residues of $H(x,y)$. Since those residues are polynomial, and because of the definition of the transformation $\mathcal{I}$, we can truncate those equations to take care only of the terms which will contribute to a given order.

Let us write $P_{k,p}(X)$ for the polynomial $P_k(X)$ truncated to a degree less or equal to~\(p\), and  $Q_{k,p}(X)$
for the polynomial $Q_k(X)$ similarly truncated. Then the equation~(\ref{equa_H}) for \(L_k\) becomes
\begin{equation}
  (k-2\gamma -3\gamma a \partial_a ) L_k=Q_{k,p-1}(\partial_{L_1}\partial_{L_2})G(L_1)G(L_2)|_{L_1=L_2=0},
\end{equation}
and the right hand side of the equation for \(\gamma\) in~(\ref{SDE_simples}) gets an additional part in its right hand side,
\begin{equation}
 a\sum_{k=2}^{+\infty}L_k +a\sum_{k=2}^{+\infty} {\I}\left(P_{k,p-1}(xy)\left[\frac{2}{k+x}-\frac{1}{k}\right]\right)
\end{equation}
The $2$ in the equation for $\gamma$ is there because for each $k$, $H(x,y)$ has a pole $x=-k$ and at $y=-k$ which give equal contributions. Moreover $P_{k,p}(X)$ is just the polynomial $P_k(X)$ truncated to a degree less or equal to~\(p\), and similarly for $Q_{k,p}(X)$. 
${\I}$ is the linear transform defined in section~\ref{methods}, which reduced in this pole expansion to: 
\begin{align}
\begin{cases}
 & {\I}\left(\frac{(xy)^n}{k+x}\right) = (-k)^n\gamma_n\left[F_k-\frac{1}{k}\sum_{i=0}^{n-1}\left(-\frac{1}{k}\right)^i\gamma_i\right], \\
 & {\I}\left((xy)^n\right) = \gamma_n^2.
\end{cases}
\end{align}
We have used:
\begin{equation*}
 \frac{(xy)^n}{k+x} = (-k)^ny^n\left[\frac{1}{x+k}-\frac{1}{k}\sum_{i=0}^{n-1}\left(-\frac{x}{k}\right)^i\right].
\end{equation*}
With this definition of ${\I}$ we see that only the term $(xy)^{p-1}$ is needed for the solution at the order $a^{2p+1}$ of \(\gamma\) since the leading term of
 $a {\I}\left((xy)^p\right)$ which is $a\gamma_n^2$ is of order $2p+1$. 
 However, when looking at the coefficients of the symbols \(A\) and~\(B\) in \(\gamma_n\), they still are proportional to \(a\). The term \( a(xy)^p \) will therefore contribute terms of order \(a^{p+2}\).

Last, but not least, the equation for $F_k$ is similar to the one for $F$, and is given in (\ref{equa_F}). This equation will never change, whatever the order one needs, simply because any new term which might affect $F_k$ will come through changes in~$\gamma$. The modifications to the $L_k$ functions instead come also from changes to the equation of $L_k$. 

\subsection{Solutions to the Schwinger--Dyson equations with the (xy) term}

To start our study of the effect of the infinitude of poles, we will take only the $(xy)$ contribution to each pole. Hence the Schwinger--Dyson equation comes from the following approximate Mellin transform:
\begin{eqnarray}
 \frac{1}{a}h(x,y) &=& (1+xy)\left(\frac{1}{1+x} + \frac{1}{1+y}-1\right) + \frac{1}{2}\frac{xy}{1-x-y}\nonumber\\
 	&& +\sum_{k=2}^{+\infty}\left(\frac{1}{k+x} + \frac{1}{k+y} -\frac{1}{k}\right) \frac{xy}{k(k-1)} + \sum_{k=2}^{+\infty}\frac{1}{k-x-y}\frac{xy}{k(k+1)}.
\end{eqnarray}
In the equation for $L_k$ we use only the linear term for the numerator in~(\ref{equa_H}):
\begin{align*}
 [k-\gamma(2+3a\partial_a)]L_k & = \frac{1}{k(k+1)}\partial_{L_1}\partial_{L_2}G(L_1)G(L_2)|_{L_1=L_2=0} \\
                               & =  \frac{1}{k(k+1)}\gamma^2
\end{align*}
For the Schwinger--Dyson equation, one has simply to apply Eq.~(\ref{SDE}). Some series arise, which are easily computable.  So we end up with five coupled non-linear partial 
differential equations to solve,
\begin{subequations}
\begin{align}
 & F = 1 - \gamma(3a\partial_a+1)F  \label{equations1} \\
 & L = \gamma^2 + \gamma(3a\partial_a+2)L \\
 & kF_k = 1 - \gamma(1+3a\partial_a)F_k \\
 & kL_k = \frac{1}{k(k+1)}\gamma^2 + \gamma(2+3a\partial_a)L_k \\
 & \gamma = 2aF - a -2a\gamma(F-1) + \frac{1}{2}aL +2a\gamma -a\gamma^2[3-\zeta(2)] + a\sum_{k=2}^{+\infty}\left(L_k-2\gamma\frac{F_k}{k-1}\right) \label{equation5}
\end{align}
\end{subequations}
with $\zeta$ being Riemann's zeta function. One may be worried by the $\zeta(2)$ in the last equation. Indeed, using the link between the logarithm 
of the Euler's gamma function, the Riemann's zeta function and the Euler--Mascheroni's constant $\gamma$:
\begin{equation*}
 \ln\Gamma(z+1) = -\gamma z+\sum_{k=2}^{+\infty}\frac{(-1)^k}{k}\zeta(k)z^k
\end{equation*}
(for example in \cite{abramowitz}), it is easy to see that one can write the exact Mellin transform as an exponential of a sum over the odd values of the Riemann's zeta 
function:
\begin{equation} \label{Hzeta}
 H(x,y) = \frac{a}{1+x+y}\exp\Bigl(2\sum_{k=1}^{+\infty}\frac{\zeta(2k+1)}{2k+1}\left((x+y)^{2k+1}-x^{2k+1}-y^{2k+1}\right)\Bigr).
\end{equation}
The Mellin transform was already written in this form in \cite{BeSc08}, so one expects only odd zeta values in the result. However, the sum will give compensating terms and this $\zeta(2)$ will not appear any more in the result. This provides a check that the calculations are correct.

Now, as in Section~\ref{FirstOrder}, we can define the functions $f_k$, $g_k$ and $h_k$, and $l_k$, $m_k$ and $n_k$ for the functions $F_k$ and $L_k$. Then the system of equations
(\ref{equations1})-(\ref{equation5}) shall be rewritten for those functions. One ends up with fifteen coupled partial non-linear differential equations for fifteen 
functions, that we will not write down explicitly. 

Solving those equations should be done in the same order than in the section~\ref{methods}, with the equations for $f_k$ and $l_k$ solved with the equations for
$f$ and $l$, and similarly those for $g_k$, $m_k$, $h_k$ and~$n_k$ with \(h\) and~\(m\). As in the previous case, the order two terms of $n(a)$ and~$g(a)$ are not fixed by the $a^2$ equations. However we only need the order three terms of the equations for \(g\) and \(n\) to fix this,  while the most tedious equations to 
solve at a given order are those for $d$ and~$e$ which involve sums over \(k\). Fixing those two last coefficients thus does not add much complexity. Moreover, we do not need to add more terms in the $\gamma$ equation, since we are looking for the equation on the coefficient $c$ of $\gamma$, and the higher order 
(such as the $(xy)^2$ term) will act on the $d$ and $e$ terms only, thanks to the relations (\ref{relation_symboles}). The already computed orders $a^0$ and $a^1$ are unchanged by the 
addition of the new terms as expected and, all computations being done, we end up with the solution to the equations (\ref{equations1})-(\ref{equation5}) up to the order $a^2$.
\begin{subequations}
\begin{align}
  & f(a) = 1-a+6a^2\\
  & g(a) = g_0\left(1+\frac{16}{3}a+\frac{2}{9}\left[-65+12\zeta(3)\right]a^2\right)\\
  & h(a) = n_0\left(-\frac{1}{4}a+\frac{29}{12}a^2\right)\\
  & l(a) = a^2\\
  & m(a) = 2a^2g_0\\
  & n(a) = n_0\left(1-\frac{11}{3}a+\frac{8}{9}\left[28-3\zeta(3)\right] a^2\right)\\
  & f_k(a) = \frac{1}{k}-\frac{1}{k^2}a+\frac{2(2+k)}{k^3}a^2 \\
  & g_k(a) = -\frac{2g_0}{k(k-1)}\left(a+\frac{12-28k+13k^2}{3k(k-1)}a^2\right) \\
  & h_k(a) = \frac{n_0}{k(k+1)}\left(-\frac{a}{2}+\frac{6+16k+7k^2}{3k(k+1)}a^2\right) \\
  & l_k(a) = \frac{a^2}{k^2(k+1)} \\
  & m_k(a) = \frac{4g_0}{k(k+1)^2}a^2 \\
  & n_k(a) = \frac{n_0}{(k-1)k(k+1)}a^2 \\
  & c(a) = 1-2a+14a^2\\
  & d(a) = 2g_0\left(1+\frac{13}{3}a+\frac{2}{9}\left[-71+12\zeta(3)\right]a^2\right)\\
  & e(a) = n_0\left(\frac{1}{2}-\frac{7}{3}a+\frac{1}{6}\left[\frac{329}{3}-8\zeta(3)\right]a^2\right)
 \end{align}
\end{subequations}
with the $n_2$ and $g_2$ being fixed by a computation at the $a^3$ order.\footnote{
We use the same notations to denote the functions \(g_k(a)\), appearing as factors of \(A\) in \(F_k\) and the coefficients \(g_i\) of the function \(g(a)\), in order to keep a strong parallelism between the expansion of \(F\) and \(F_k\), and similarly for \(n_k(a)\) and~\(n_i\). We hope that the context make the two different usages clear.
}
\begin{align*}
 & g_2 = \frac{2}{9}\left[-65+12\zeta(3)\right]g_0 \\
 & n_2 = \frac{8}{9}\left[28-3\zeta(3)\right]n_0
\end{align*}
So this order is a nice check of our procedure since the two first order are unchanged and $\zeta(2)$ disappears everywhere as expected. Now, we feel confident about the method and we will reach the fourth order.

\section{Two more orders}

\subsection{The $a^3$ order}

We first have to determine the coefficients of \(P_{k,2}(X)\), which appears in the equation for $\gamma$. It is simply:
\begin{equation}
 P_{k,2}(X) = \frac X {k(k-1)} +  \frac{X^2}{k^2(k-1)}\left(H_k+H_{k-2}\right).
\end{equation}
This is true for all values of $k$ with the convention that $H_k$, the $k^{th}$ harmonic number, is defined by $H_0=0$, $H_{k}=H_{k-1}+1/k$. Then the equation for $\gamma$ becomes:
\begin{align}
 \gamma & = 2aF - a -2a\gamma(F-1) + \frac{1}{2}aL  + a\sum_{k=2}^{+\infty}\left(L_k-2\gamma\frac{F_k}{k-1} + 2\gamma_2F_k\frac{H_k+H_{k-2}}{k-1}\right) +2a\gamma \nonumber \\
\llcorner &  -a\gamma^2[3-\zeta(2)] -6a \gamma_2+  2a\gamma_2\gamma\bigl[6-\zeta(2)-3\zeta(3)\bigr] - a\gamma_2^2 
	\bigl [10-2\zeta(2)-\textstyle{\frac{3}{2}}\zeta(4)-4\zeta(3)\bigr].
\end{align}
The only other equation to be changed is the one for $L_k$ which gets a new term
\begin{equation*}
 Q_{k,2}(X)=\frac1{k(k+1)}X -\frac{2H_{k-1}}{k^2(k+1)}X^2
\end{equation*}
and so the equation for $L_k$ is now:
\begin{equation}
 kL_k = \gamma(3a\partial_a+2)L_k+\frac{1}{k(k+1)} \gamma^2 -\frac{2H_{k-1}}{k^2(k+1)}\gamma_2^2.
\end{equation}
These equations (together with the three equations for the other functions) can now be solved at the third order. For the sake of readability, we will not write the fifteen functions at this 
order, but only the functions which are a part of $\gamma$. One ends up with a solution without any $\zeta(2n)$,
\begin{subequations}
\begin{align} \label{ordre3}
  & c(a) = 1-2a+14a^2+16\left[\zeta(3)-10\right]a^3 \\
  & d(a) = g_0\left(2+\frac{26}{3}a+\left[-\frac{284}{9}+\frac{16}{3}\zeta(3)\right]a^2+\frac{4}{9}\left[\frac{7873}{9}-134\zeta(3)\right]a^3\right) \\
  & e(a) = n_0\left(\frac{1}{2}-\frac{7}{3}a+\frac{1}{6}\left[\frac{329}{3}-8\zeta(3)\right]a^2+\frac{1}{9}\left[-\frac{33889}{18}+188\zeta(3)\right]a^3\right)
\end{align}
\end{subequations}
We have already included the  coefficients $g_3$ and $n_3$.
\begin{align*}
 & g_3 = g_0\frac{8}{9}\left[\frac{1687}{9}-8\zeta(3)\right] \\
 & n_3 = \frac{n_0}{81}\left[-22207+3168\zeta(3)\right]
\end{align*}
Again, the disappearance of every even zeta values from the final result is a very useful fact. It provides us a check of our computations.

\subsection{The $a^4$ order}

For the $a^4$ order, we need to compute the coefficient of degree two in a product of linear terms. We use:
\begin{equation*}
 \prod_{i=1}^n\left(1+\alpha_iX\right) = 1+X\sum_{i=1}^n\alpha_i+\frac{X^2}{2}\left(\Bigl[\sum_{i=1}^n\alpha_i\Bigr]^2-\sum_{i=1}^n\alpha_i^2\right)+\mathcal{O}(X^3).
\end{equation*}
One finds easily the coefficient of the cubic term of the $P_k(X)$ polynomials (\ref{residusPk})
\begin{equation} \label{ordre3}
	p_{k,3} =  \frac{1}{k^3(k-1)}\biggl(H_kH_{k-2}+\sum_{1\leq i<j\leq k}\frac{1}{ij}+\sum_{1\leq i<j\leq k-2}\frac{1}{ij}\biggr)
\end{equation}
The last sum is not defined for $k=2$ and $k=3$ and we will  write those two cases separately, with the values
\begin{align*}
  p_{2,3} &= \frac{1}{16} \\ 
  p_{3,3} &= \frac{17}{324}
\end{align*}
This phenomenon of a general term undefined for the first coefficients will appear for the coefficient of any term $(xy)^n$, and we would have to deal with it at any further order.
\footnote{However, one might notice that we find the right values of the cubic terms of those two first polynomials if we simply set to zero the undefined term in (\ref{ordre3}). Since we don't 
have a proof of this effect being true at any order, it seemed to be simpler to separate the first terms off the others.}

Now, using
\begin{equation}
  \frac{(xy)^3}{k+x} = -k^3y^3\left(k\frac{1}{k+x}-1+\frac{x}{k}-\frac{x^2}{k^2}\right)  \longrightarrow -k^2\gamma_3\left(kF_k-1+\frac{\gamma}{k}-\frac{\gamma_2}{k^2}\right)
\end{equation}
we end up with the following equation for $\gamma$:
\begin{equation}
 \gamma=\text{[orders 0, 1, 2 and 3]}+a\gamma_3\sum_{k=2}^{+\infty}\left[-2k^3F_k + 2k^2 - 2k\gamma + 2\gamma_2 - \frac{\gamma_3}{k}\right]p_{k,3}+a\sum_{k=2}^{+\infty}L_k.
\end{equation}
We can use (\ref{ordre3}) in there and the values of the \(p_{k,3}\) in this equation. Many series will arise, which could all be computed in terms of zetas, multizetas and rational numbers. These computations have some interesting 
features, justifying working them out. However, such a computation is complex and it is a better strategy to not separately sum each series, but rather to combine the generic terms of the series. We need the following expansion:
\begin{equation*}
 -2k^3F_k + 2k^2 - 2k\gamma + 2\gamma_2 - \frac{\gamma_3}{k} = S_k + T_kA + U_kB.
\end{equation*}
The series $S_k$, $T_k$ and $U_k$ are not so complicated anymore, since we only need their dominant terms:
\begin{subequations}
\begin{align}
  & S_k = \frac{28}{k}a^3+\mathcal{O}(a^4) \\ 
  & T_k = ag_0\left[4\frac{k^2}{k-1}-4k-4-\frac{2}{k}\right]+\mathcal{O}(a^2) = ag_0\left[\frac{4}{k-1}-\frac{2}{k}\right]+\mathcal{O}(a^2) \\
  & U_k = an_0\left[\frac{k^2}{k+1}-k+1-\frac{1}{2k}\right]+\mathcal{O}(a^2) = an_0\left[\frac{1}{k+1}-\frac{1}{2k}\right]+\mathcal{O}(a^2)
\end{align}
\end{subequations}
The higher order terms are not needed here since there is a $\gamma_3$ in front of those sums which starts at the $a^3$ order for its low order part and at the $a^1$  one for the other parts:
\begin{equation*}
 \gamma_3 = 28a^3+\mathcal{O}(a^4) + 2g_0A\left((a+\mathcal{O}(a^2)\right)+\frac{1}{2}n_0B\left(a+\mathcal{O}(a^2)\right).
\end{equation*}
These results only depend on the lowest order values of \(\gamma\) together with the renormalisation group equation (\ref{recursion_gamma}) and the relations between the symbols $A$, $B$, 
and their derivatives (\ref{relation_symboles}).

The contribution of the $(xy)^3$ term for $c_4$ is vanishes since there is no $a^4$ term without $A$ and $B$. We are simply left with the following series:
\begin{align}
 R_1 & = a\sum_{k=2}^{+\infty}\left[2g_0S_k+28a^3T_k\right]p_{k,3} \nonumber \\
     & = 112g_0a^4\left[\frac{115}{1296}+\sum_{k=4}^{+\infty}\frac{p_{k,3}}{k-1}\right] \\
 R_2 & = a\sum_{k=2}^{+\infty}\left[\frac{1}{2}n_0S_k+28a^3U_k\right]p_{k,3}\nonumber \\
     & = 28n_0a^4\left[\frac{71}{2592}+\sum_{k=4}^{+\infty}\frac{p_{k,3}}{k+1}\right].
\end{align}
Hence we got $\gamma_3\sum_{k=2}^{+\infty}\left[-2k^3F_k + 2k^2 - 2k\gamma + 2\gamma_2 - \frac{\gamma_3}{k}\right]p_{k,3}=R_1A+R_2B+\mathcal{O}(a^5)$. Those sums are still not very simple, but 
much simpler than the ones we had before. The Schwinger--Dyson equations are now written in a very compact form:
\begin{equation} \label{SDE_fin}
 \gamma=\text{[orders 0, 1, 2 and 3]}+a(R_1A+R_2B)+a\sum_{k=2}^{+\infty}L_k
\end{equation}
One still has to add the $(xy)^3$ term into the equation of $L_k$, which depends on the coefficient cubic \(q_{k,3}\) of \(Q_k(X)\):
\begin{equation}
 q_{k,3}=\frac{1}{k^3(k+1)}\left(2H_{k-1}^2-H_{k-1,2}-2\frac{H_{k-1}}{k}\right).
\end{equation}
Here, $H_{k,n}$ denote the generalized harmonic numbers defined by $H_{0,n} = 0$ and $H_{k,n} = H_{k-1,n} + 1/k^n$.
Hence, at this order, the equation for $L_k$ becomes:
\begin{equation}
 kL_k = \gamma(3a\partial_a+2)L_k+\frac{1}{k(k+1)}\gamma^2-\frac{2H_{k-1}}{k^2(k+1)}\gamma_2^2
 +q_{k,3}\left(\gamma_3\right)^2.
\end{equation}
The equations for $F$, $L$ and $F_k$ are unchanged and we end up, after these simplifications, with a system of five coupled equations.
\begin{subequations}
\begin{align} \label{equations12}
 & F = 1 - \gamma(3a\partial_a+1)F \\
 & L = \gamma^2 + \gamma(3a\partial_a+2)L \\
 & kF_k = 1 - (1+3a\partial_a)F_k \\
 & kL_k = \gamma(3a\partial_a+2)L_k+\gamma^2\frac{1}{k(k+1)}-\frac{2H_{k-1}}{k^2(k+1)}\gamma_2^2
 	+q_{k,3}\left(\gamma_3\right)^2\\
 & \gamma=\text{[orders 0, 1, 2 and 3]}+a(R_1A+R_2B)+a\sum_{k=2}^{+\infty}L_k \label{equation52}
\end{align}
\end{subequations}
We can solve them in order to get the anomalous dimension of the massless Wess--Zumino model up to the fourth order. For the sake of readability, we will write only this fourth order:
\begin{subequations}
 \begin{align}
  & c_4 = 2444-328\zeta(3) \label{resultat41} \\
  & d_4 = 2g_4 + \frac{1}{81}g_0\left[-73720 +65952\zeta(3)\right] \\ 
  & e_4 = \frac{1}{2}n_4 + n_0\left[\frac{384227}{324}-\frac{974}{9}\zeta(3)\right] \label{resultat42}
 \end{align} 
\end{subequations}
One striking observation about our result (\ref{resultat41}) - (\ref{resultat42}) is that they contain only rational numbers and $\zeta(3)$, when the summation over \(k\) gives multizetas of weight 5 for \(d_4\) and~\(e_4\). However the highest weight terms cancel each others, so that the weight is not higher than the one for \(c_4\), where the \((xy)^3\) does not contribute and every sum is of weight smaller than 4.

Now, let us look at the final solution. The coefficients $g_4$ and $n_4$ could be fixed by going to the fifth order, thanks to the equations of $F$ and $L$ respectively.
\begin{subequations}
 \begin{align}
  & g_4 = g_0\left[-\frac{652516}{243} + \frac{5212}{27}\zeta(3) + 168\zeta(5) + \frac{32}{9}\zeta(3)^2\right] \\
  & n_4 = n_0\left[\frac{3741119}{1944} - \frac{8522}{27}\zeta(3) - 84\zeta(5) + \frac{16}{9}\zeta(3)^2\right]
 \end{align}
\end{subequations}
So we end up with the final values for the fourth order of the anomalous dimension of the massless Wess--Zumino model.
\begin{subequations}
 \begin{align} \label{sol_a4}
  & c_4 = 2444-328\zeta(3) \\
  & d_4 = -g_0\left[\frac{1526192}{243} +\frac{32408}{27}\zeta(3)+ 336\zeta(5) + \frac{64}{9}\zeta(3)^2\right] \\ 
  & e_4 = n_0\left[\frac{6046481}{1944}-\frac{11444}{27}\zeta(3)-84\zeta(5)+\frac{16}{9}\zeta(3)^2\right] \label{sol_a4_fin}
 \end{align} 
\end{subequations}
So $d_4$ and $e_4$ finally involve some $\zeta(5)$ and $\zeta(3)^2$. This highest weight terms stem from the order~5 in the equation for \(g(a)\) (resp.\ \(n(a)\)), which contains \(\zeta(5)\) 
through \(g_0 c_5\) (resp.\ \(n_0 c_5\)) and \(\zeta(3)^2\) through \(g_2 c_3\) (resp.\ \(n_2 c_3\)).  The similarity of their origin explains that this highest weight terms differ simply by a factor \(\pm4\). 

Now, let us go back to the weight of the zetas in the low order part. We want to prove that the weight in \(\zeta\) of~\(c_p\), which is the coefficient of \(a^{p+1}\) in~\(\gamma\), is \(p\) or less. If we suppose  this property true, the renormalisation group equation~(\ref{recursion_gamma}) allows to show that the coefficient of \(a^{n+p}\) in $\gamma_n$ is of maximal weight \(p\). 
From the expression (\ref{Hzeta}) of $H(x,y)$, it is clear that the derivative of total order \(k\) of \(H(x,y)\) has maximal weight \(k\). The same upper bound on the weight can be deduced from its expression as a sum over the poles, but it is however highly non trivial in this case that only products of zeta values at odd integers appear. It then follows that for every term \(h_{n,m}\gamma_n\gamma_m\) in Eq.~(\ref{SDE}), the terms of degree \(p+1\) in \(a\) is of maximal weight less or equal to \(p\). Our hypothesis on the weight of the zetas appearing in \(\gamma\) can therefore be proved by induction.

Now, for the parts of $\gamma$ proportional to $A$ and $B$ the reasoning made for the low order part does not hold. The highest weight terms in the sums over the poles cancel in the terms we have studied, so that we do not have the weights \(2p\) for the coefficient of \(a^p\).  In fact, it seems that the weight of the \(p\)-th correction is still \(p\), but with a modified weight, where \(\zeta(2r+1)\) is only of weight~\(2r\).

\section{Comparison to numerical results}

Now that we have this nice analytical solution (\ref{sol_a4})-(\ref{sol_a4_fin}) and the lower orders of the Schwinger--Dyson equation (\ref{SDE}) we will compare our results with the numerical solution obtained for
\cite{BeSc08}. 

The objective of this study is twofold.
On one hand, we want to find $g_0$ and $n_0$ since they remain as free parameters in our analytical study. We will fit $g_0$ and $n_0$ to make our computed asymptotic behavior match the data of \cite{BeSc08} at two consecutive orders. On the other hand, we are 
expecting the convergence of the fit to become faster when we take into account higher orders of our solution. So looking for speed of convergence and how it changes when we include 
higher orders will be a check of our computations.

The obtained values of $g_0$ are presented in figure~\ref{fig_g0} as a function of the order at which the fit is done for different approximations of the asymptotic behavior. Likewise, the 
values for \(n_0\) are presented in figure~\ref{fig_n0}.
\begin{figure}[t]
\caption{$2g_0$ from different fits.} \label{fig_g0}
 \includegraphics[scale=0.5]{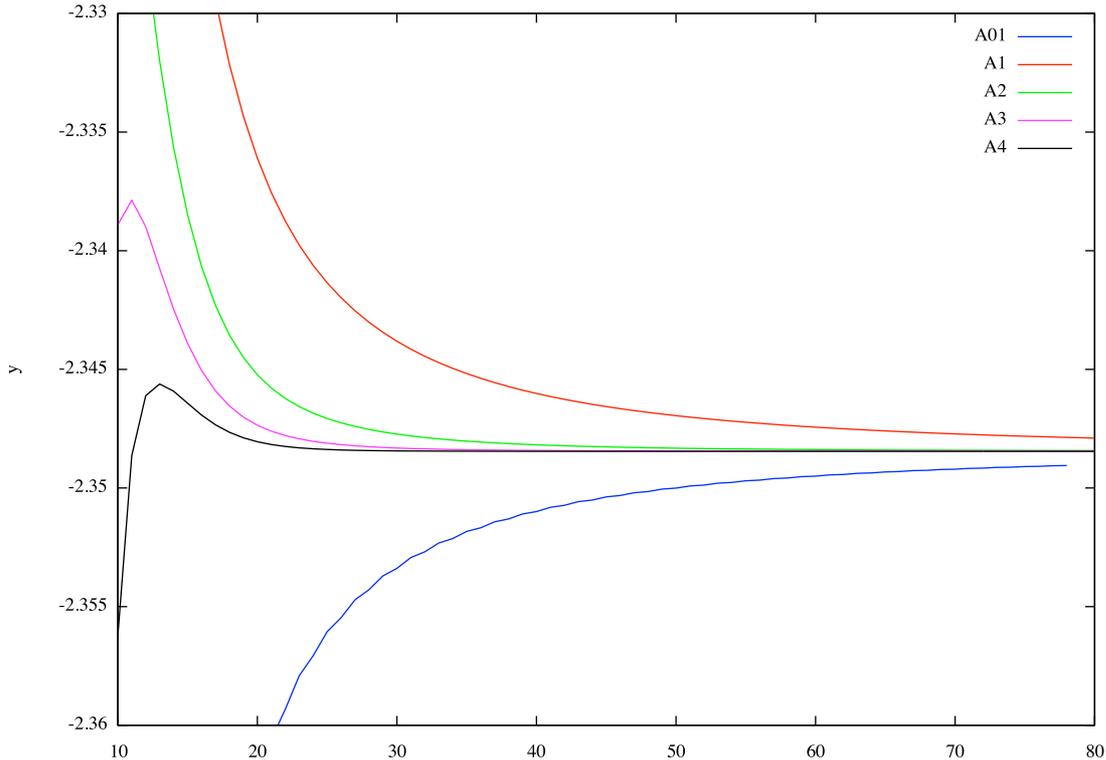}
\end{figure}
\begin{figure}
\caption{$\frac{1}{2}n_0$ from different fits.} \label{fig_n0}
 \includegraphics[scale=0.5]{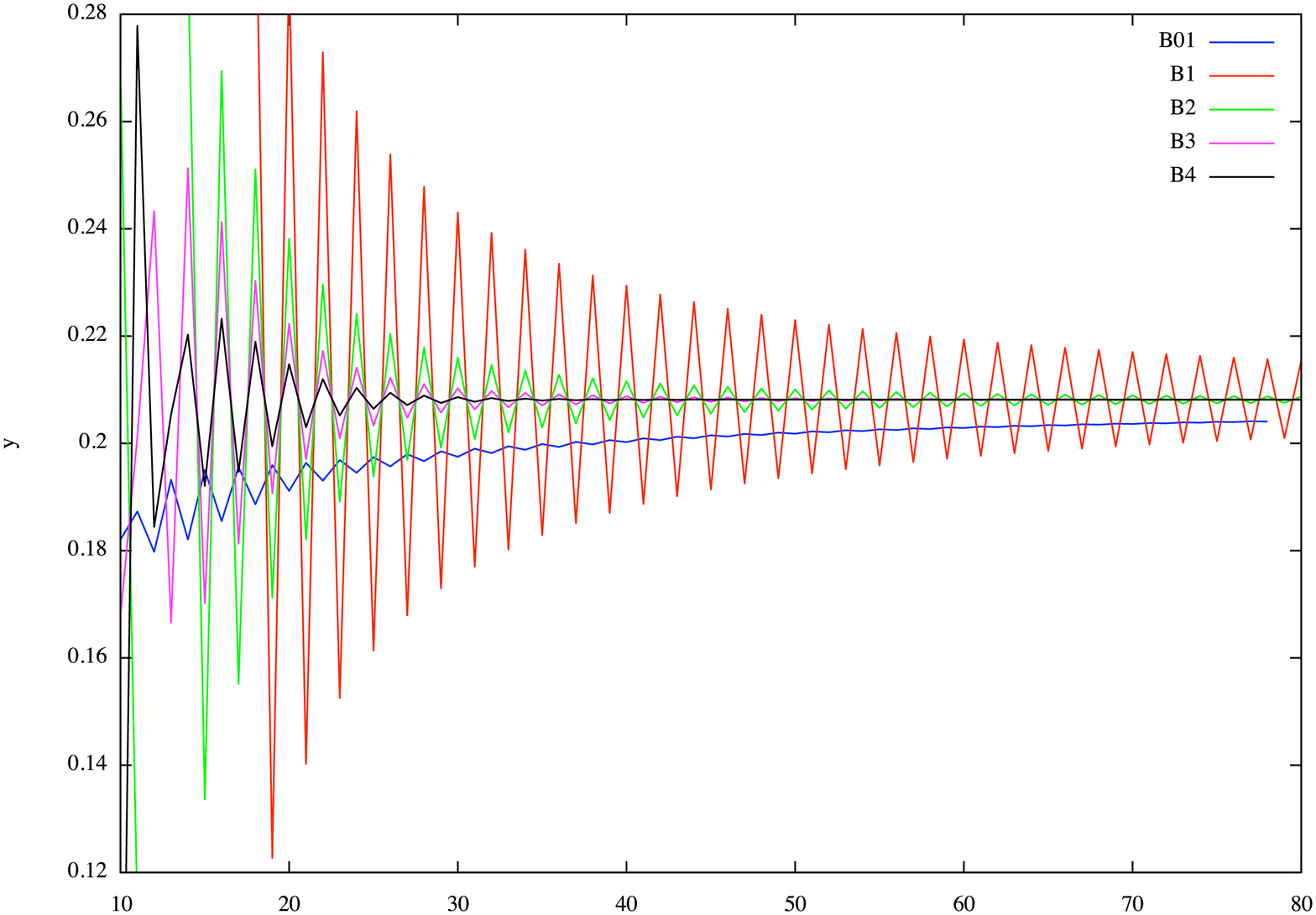}
\end{figure}
The curves $Ai$ or $Bi$ are obtained when one approximates the asymptotic behavior by including terms of up to order \(i\) in $d(a)$ and~$e(a)$. Since the case without any correction has very 
poor convergence, we plot $A01$ and $B01$, which correspond to fits on three values on a combination of \(A\), \(B\) and~\(aA\), without imposing the relation we deduced between the two terms 
proportional to \(A\). One clearly see the convergence improvement when using more terms of the asymptotic series. This can be seen as a check of our computations by numerical experiment.

We numerically get $2g_0\simeq-2.3484556$ and $\frac{1}{2}n_0\simeq0.208143(4)$.   The relative precision is better on $g_0$ than on \(n_0\), which was expected since the $A_n$ sequence growths 
faster than the $B_n$ one. To improve the precision on \(g_0\) and \(n_0\), one can either go to higher order in \(a\) or compute additional terms in the asymptotic expansion.   Had we not have the numerical results of~\cite{BeSc08}, we probably could obtain the same precision on \(g_0\) and \(n_0\) with fewer low order terms of \(\gamma\) and some additional terms of the asymptotic 
behavior, for a smaller total computational cost. This is not so important here where computations remain manageable, but could be of serious interest when adding higher loop corrections to the Schwinger--Dyson equation.

\section*{Conclusion}

In this work we have gone further on the path defined in \cite{Be10a}.  The contributions of all the poles of the Mellin transform could be worked out, allowing us first to recover the first orders of the perturbative solution, but also to reach some non-perturbative information about a QFT without too many heavy computations.

At the level of the nature of the coefficients obtained, we observe that the sum over the poles does not produce zeta values with weights growing as twice the order, as could be expected. The final result seems compatible with a higher weight of the coefficients of \(d_n\) or \(e_n\) of \(n\) in a system where \(\zeta(2p+1)\) has weight \(2p\). Uncovering the precise mechanisms through which the highest weight terms cancel each others, how all the multiple zeta values stemming from the different sums hopefully conspire to give only products of \(\zeta\)-values for odd integers looks like a combinatorial nightmare in search of a conceptual solution.

We have been able to compute the $a^5$ order of the $\gamma$ function around $0$. The dominant asymptotic behaviors of the perturbative series encoded by  $A$ and $B$ give rise to the first two singularities of the Borel transform of $\gamma$. 
These singularities are $\left(\xi + \frac{1}{3}\right)^{-5/3}$ from $A$ and $\ln\left(\xi-\frac{1}{3}\right)$ from $B$. Multiplying a function by \(a\) corresponds to taking a primitive of its Borel transform, so that the terms of \(d\) and~\(e\) we computed give the first derivatives of an analytic function multiplying these basic singularities to give the full singularity of the Borel transform at the point \(-1/3\) and \(+1/3\).

We have compared our analytical results to numerical ones and find an excellent agreement between them. Moreover, this study allowed us to numerically determine the two last unconstrained 
parameters. Such a control on the singularities of the Borel transform is of great interest for the definition of the renormalization group function \(\beta\) through a Borel sum. There may be however additional singularities of the Borel transform. 
In fact, similar computations should allow the determination of the other possible singularities of the Borel transform, but the overall factors should be much more difficult to obtain, since this would entail a determination of the analytic extension of the Borel transform beyond its convergence disk.

An essential ingredient of our computations has been the possibility of expressing the Mellin transform of the diagram as a sum over poles, which allowed us to sum up infinite numbers of terms in its Taylor expansion which have contributions of the same order to the asymptotic behavior. 
However, the way in which the residues are extended to all values of the variables is not fully justified: the sum over the poles becomes convergent and defines a meromorphic function. Thus the true Mellin transform can differs from this meromorphic function only by an entire function which vanishes at the origin together with an infinite number of its derivatives. We have no proof that it is exactly zero. Such a result would be interesting on many accounts: it would give a firmer ground to our computations, it would imply an infinite number of relations 
between multizeta values through the comparison of the expansion in poles and the one in terms of \(\zeta\) values at odd integers~(\ref{Hzeta}). Finally, should it generalize to higher loop 
graphs, it would be a very interesting way of evaluating  Mellin transforms of multi loop graphs, since residues are also in these cases explicit polynomials.

In this work, we have only been up to the order $a^5$ in $\gamma$. There is no reason not to go beyond, but the increasing technicalities of the computation. Moreover, there is a lack of a full table of relations among multi zeta values which makes the calculations quite lengthy. Such a difficulty might be overturned by a careful combinatorial analysis of the quantities arising in our problems, and is left for further investigation.

An other interesting trail to follow would be the study of higher loop corrections of the Schwinger--Dyson equation, furthering the study of~\cite{BeSc12}. Indeed the next term is a three loop term, which therefore modifies the coefficient \(c_2\), and through it, the coefficients \(d_1\) and \(e_1\). 
 Even a large \(N\) limit would involve a four loop primitive divergence, so that this analysis must be complete to take into
account those higher loop terms. The fact that our method has a fairly low computational cost will be very interesting in this next step. Indeed with the three (or more) loops terms
in the Schwinger--Dyson equation, the number of propagators will increase rapidly, but the number of poles with differing contributions in our method does not grow so fast.

Moreover, we have worked in a model where the only relevant Schwinger--Dyson equation for the renormalization group is only for the propagator. In the more physically relevant case of Yang--Mills theories, one also has to deal with 
Schwinger--Dyson equation for vertices, and massive particles. But this involves many new difficulties that we hope to address one by one.

\bibliographystyle{unsrt}
\bibliography{renorm}

\end{document}